\documentstyle[11pt,newpasp,twoside,epsf,natbib,url]{article}
\def\msun{M$_{\sun}$}

\def\edcomment#1{\iffalse\marginpar{\raggedright\sl#1\/}\else\relax\fi}
\reversemarginpar

\begin{document}

\title{A 20-cm Survey for Pulsars in Globular Clusters using the GBT
  and Arecibo}

\author{Scott Ransom$^{1,2}$, Jason Hessels$^1$, Ingrid Stairs$^3$,
  Victoria Kaspi$^{1,2}$, Paulo Freire$^4$, \& Donald Backer$^5$}

\affil{$^1$Dept. of Physics, McGill University, 3600 University St.,
  Montreal, QC H3A~2T8, Canada}
\affil{$^2$Center for Space Research,
  Massachusetts Institute of Technology, Cambridge, MA 02139}
\affil{$^3$Dept. of Physics and Astronomy, University of British
  Columbia, 6224 Agricultural Road, Vancouver, BC V6T~1Z1, Canada}
\affil{$^4$National Astronomy and Ionosphere Center, Arecibo
  Observatory, HC~03 Box 53995, Arecibo, PR 00612}
\affil{$^5$Dept. of Astronomy and Radio Astronomy Laboratory,
  University of California at Berkeley, 601 Campbell Hall 3411,
  Berkeley, CA 94720}

\begin{abstract}
  We have been conducting deep searches at $\sim$20\,cm of $>$30
  globular clusters (GCs) using the 305-m Arecibo telescope in Puerto
  Rico and the 100-m Green Bank telescope (GBT) in West Virginia.
  With roughly 80\% of our search data analyzed, we have confirmed 13
  new millisecond pulsars (MSPs), 12 of which are in binary systems,
  and at least three of which eclipse.  We currently have timing
  solutions for five of these systems and basic orbital and spin
  parameters for six others.
\end{abstract}

\section{Introduction}

The number of known MSPs in GCs has increased significantly in the
last few years due to a number of targeted surveys (see the review by
F.~Camilo and the contributions by A.~Possenti and B.~Jacoby in this
volume) that have benefitted from increased computational resources,
new large-bandwidth pulsar backends (primarily at 20\,cm, and improved
search algorithms.  Approximately half of the currently known $\sim$80
pulsars in 23
GCs\footnote{\url{http://www.naic.edu/~pfreire/GCpsr.html}} were found
in the past four years.  During the last three years we have been
searching more than 30 GCs with the Arecibo and Green Bank telescopes.
The high time and frequency resolution of these data, along with newly
developed search algorithms \citep*{rem02,rce03}, makes us
significantly more sensitive than past surveys to sub-millisecond
pulsations as well as to pulsars in ultra-compact binary systems.  To
date, we have confirmed 13 new MSPs in six GCs.

\section{Observations}

\subsection{Arecibo}

We observed all 22 GCs visible from Arecibo and within 50\,kpc of the
Sun using the L-Wide receiver (1.1$-$1.7\,GHz) and the Wideband
Arecibo Pulsar Processors (WAPPs).  The WAPPs are digital correlators
with adjustable bandwidth, sampling time, and number of lags
\citep*{dsh00}.  Our standard configuration for each WAPP uses
64\,$\mu$s sampling (16-bit samples) and 256 lags across 100\,MHz of
bandwidth.  The clusters were observed with a single WAPP for the full
time they are visible by Arecibo, which can be up to
$\sim$2.75\,hours.

We have been timing all our Arecibo discoveries on a roughly monthly
basis with multiple WAPPs and have also searched this data for pulsars
that may appear only occasionally when favorable scintillation causes
them to brighten.  Because of persistent and strong RFI from
$\sim$1220$-$1360\,MHz, we have generally used three WAPPs centered at
1170, 1420, and 1520\,MHz so that the data we obtain is relatively
free of RFI and suitable for searching.

\subsection{Green Bank Telescope}

In 2001 September and October, during some of the first scientific
observations taken with the GBT, we observed 12 globular clusters for
either 4\,hrs (M2, M4, M75, M80, M92, and NGC~6342) or 8\,hrs (M3, M13,
M15, M30, M79, and Pal1) at L-band using one or two Berkeley-Caltech
Pulsar Machines\footnote{\url{http://www.gb.nrao.edu/~dbacker}}
\citep{bdz+97}.  In general, the data consisted of 96$\times$1.4\,MHz
channels of 2 summed polarizations centered at 1375\,MHz and 4-bit
sampled every 50\,$\mu$s.  Large quantities of both persistent and
transient broadband interference and the very strong Lynchburg airport
radar with a period of approximately 12\,s have made data analysis
very difficult.

\subsection{Data Analysis}

We have processed each observation using the {\tt PRESTO} analysis
package \citep{ran01} using a computationally intensive multi-step
procedure on a 52-node Linux cluster at McGill called ``The Borg''.
We break each observation into small chunks of time ($\sim$20\,s) and
examine each frequency channel in both the time and frequency domains
in order to identify (and then mask out) strong interference.  The
observations are then de-dispersed, Fourier Transformed, pruned of
known periodic interference, and then searched using both
Fourier-domain acceleration searches and phase-modulation searches.
We use acceleration searches on the observations as a whole and on
various duration segments thereof.  These acceleration searches are
conducted on all clusters whether the dispersion measure is known
towards it or not.  Finally, we fold and examine by eye all candidates
above $\sim$6\,$\sigma$ that are not associated with known RFI
sources.  Additional details of the analysis procedure can be found in
\citet{rsb+04}.

\section{Results}

With $\sim$80\% of our data analyzed to our satisfaction, we have
discovered and confirmed at least 13 new MSPs, 12 of which are in
binary systems.  In addition, we have at least 2 additional binary MSP
candidates which we believe are real, but have not (as of yet)
confirmed.  The new systems include the first pulsars discovered in
four clusters: M3, M30, M71, and NGC~6749, as well as three new
pulsars each in the GCs M5 and M13.  In this section we briefly
discuss the interesting properties of the new systems, organized by
cluster.  All cluster properties come from the GC catalog of
\citet{har96}\footnote{\url{http://www.physics.mcmaster.ca/resources/globular.html}}.

\begin{table}
\caption{Newly Discovered Millisecond Pulsars in Globular Clusters}
\begin{tabular}{lcccccc}
\hline \hline
        &           & $P_{psr}$ & DM          & $P_{orb}$ & $x^{a}$ & Min $M_2^b$  \\
Pulsar  & Telescope & (ms)      & (pc/cm$^3$) & (hr)      & (lt-s)  & (M$_{\sun}$) \\
\hline \hline
M3A         & Arecibo   & 2.545 & 26.5 & Unk.        & Unk.  & Unk.       \\
M3B         & Arecibo   & 2.390 & 26.2 & 34.0        & 1.9   & 0.20       \\
M3C$^f$     & GBT       & 2.166 & 26.5 & Unk.        & Unk.  & Unk.       \\
M3D         & Arecibo   & 5.443 & 26.3 & $>$50 days  & Unk.  & Unk.       \\
M5C$^{c,e}$ & Arecibo   & 2.484 & 29.3 & 2.08        & 0.057 & 0.038      \\
M5D         & Arecibo   & 2.988 & 29.3 & 29.3        & 1.6   & 0.19       \\
M5E         & Arecibo   & 3.182 & 29.3 & 26.3        & 1.2   & 0.14       \\
M13C$^{c,d}$& GBT$+$AO  & 3.722 & 30.1 & $-$         & $-$   & $-$        \\
M13D$^c$    & Arecibo   & 3.118 & 30.6 & 14.2        & 0.92  & 0.18       \\
M13E$^{e?}$ & Arecibo   & 2.487 & 30.3 & 5.12        & 0.17  & 0.061      \\
M30A$^{c,e}$& GBT       & 11.02 & 25.1 & 4.18        & 0.23  & 0.10       \\
M30B        & GBT       & 12.99 & 25.1 & $>$15       & Unk.  & $\sim$0.35 \\
M71A$^{c,e}$& Arecibo   & 4.889 & 117  & 4.24        & 0.078 & 0.032      \\
NGC6749A    & Arecibo   & 3.193 & 194  & $\sim$day   & Unk.  & Unk.       \\
NGC6749B$^f$& Arecibo   & 4.968 & 192  & Unk.        & Unk.  & Unk.       \\
\hline \hline
\end{tabular}

$^a$ $x\equiv a_1 \sin (i)/c$. $^b$ Assuming a pulsar mass ($M_1$) of 1.4\,M$_{\sun}$. \\
$^c$ Pulsar has a timing solution.  $^d$ Isolated.  $^e$ Eclipsing.  $^f$ To be confirmed.
\end{table}

\subsection{M3}
M3 is a relatively normal cluster at a distance of $\sim$10.4\,kpc.
We have confirmed three new MSPs in M3: the 2.54-ms binary M3A, the
2.39-ms binary M3B, and the long period binary M3D, with a spin period
of 5.44\,ms.  These are the first pulsars discovered in this cluster,
and we currently have one good, but still unconfirmed, candidate, M3C
(2.16\,ms).  None of these pulsars have an average flux density large
enough to be consistently detectable with Arecibo and were all
discovered during favorable periods of scintillation. We have
sufficiently many detections of M3B for an orbital solution, and are
nearing an orbital solution for M3D (assuming that its orbit is
circular).  We have detected M3A only three times, making it currently
impossible to unambiguously determine its orbit.  We note that even
though we have not identified M3A in most of the observations, it is
likely that intense blind folding searches near its nominal pulse
period will uncover additional detections.  The M3 pulsars will be
presented in Hessels et al., in preparation.

\subsection{M5}
\begin{figure}
  \plotone{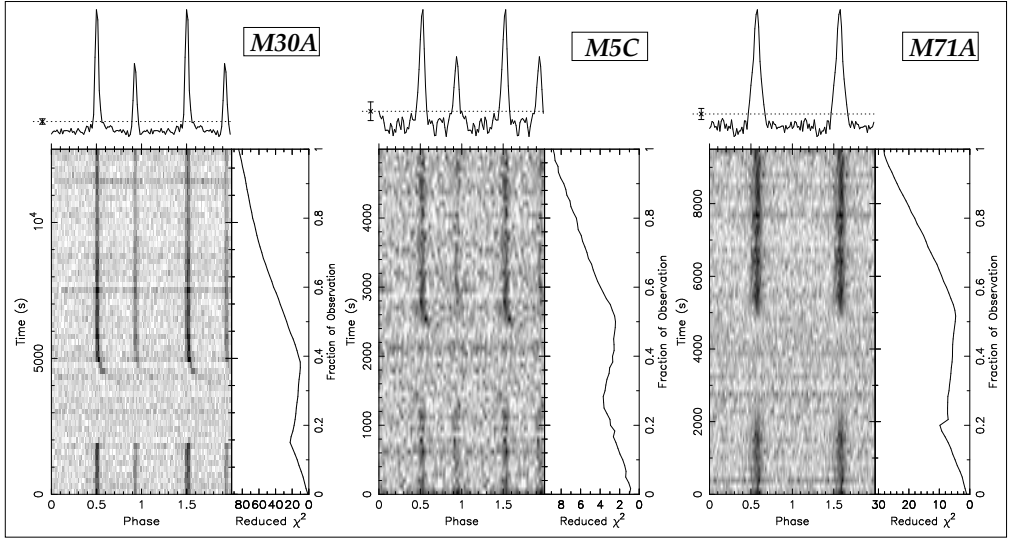}
\caption{Profiles of the three eclipsing MSPs found in the survey.  Two
cycles are shown for clarity.  The grey-scale shows the pulses as a 
function of time during the observation.  In each case the eclipse is
 clearly visible and for M30A and M5C an eclipse delay is also visible
at eclipse egress.\label{fig:eclipsers}}
\end{figure}

M5 is a fairly normal cluster at a distance of $\sim$7.5\,kpc.
\citet{awkp97} previously reported the discovery and timing of two
pulsars in the cluster: the isolated 5.5-ms pulsar M5A and the 7.9-ms
binary pulsar M5B.  Our observations quickly uncovered an eclipsing
2.48-ms ``Black-Widow'' pulsar M5C in a compact 2.1\,hr orbit (see
fig.~\ref{fig:eclipsers}).  M5C exhibits pulse arrival delays at
eclipse ingress and egress by $\ga$0.1\,ms, significantly larger than
most of the other Black-Widow pulsars.  We have also identified M5C in
an archival {\em Chandra} X-ray image.  Recently, we have uncovered
two additional MSPs in the cluster: M5D and M5E both seem to be
``normal'' binary MSPs with $\sim$0.2\,\msun\ companions in circular
$\sim$1\,day orbits.  Both pulsars are extremely faint at 20\,cm,
though, so timing solutions for the pulsars may be very difficult to
obtain.  More details will soon be available in Stairs et al., in
preparation.

\subsection{M13}
M13 is a low mass and low density cluster at a distance of
$\sim$7.7\,kpc.  \citet{and92} reported the discovery of the isolated
10.4-ms pulsar M13A and the 3.5-ms binary M13B and argued that their
presence in such a low density cluster required additional formation
mechanisms besides two-body tidal encounters.  Our observations have
uncovered 3 additional MSPs: the isolated 3.72-ms M13C, the ``normal''
binary MSP 3.12-ms M13D, and the possibly eclipsing 2.49-ms M13E.
M13E seems to lie in or near the ``gap'' splitting the Black-Widow
type binaries and the more massive Ter5A-like systems (see the
contribution by P.~Freire in this volume).  Timing measurements of
pulsars M13A$-$D show them all to be old ($\tau_{\rm c} > 2 \times
10^9$ yr) with very low surface magnetic fields (B $< 7 \times 10^8$
G).  Calculations similar to those in \citet{phi92b} show that the
measured spin-up of M13D constrains the mass-to-light ratio in the
core of the cluster to be $>$ 2 M$_{\odot}$/L$_{\odot}$ (Ransom et
al., in prep.).

\subsection{M30}
M30 is a core-collapsed cluster at a distance of $\sim$8\,kpc.  Using
the GBT, we discovered two MSPs in the cluster, one of which
(PSR~J2140-2310A) now has a full timing solution.  M30A is an
eclipsing 11-ms MSP in a 4.2\,hr orbit where the pulsed emission is
eclipsed $\sim$20\% of the time (see fig.~\ref{fig:eclipsers}).
Arrival time delays of duration up to 1$-$2\,ms are visible during
eclipse ingress and egress when scintillation (which is very strong
towards the cluster) is favorable.  M30A has likely been detected in
X-rays by {\em Chandra} and possibly in the optical by {\em HST}.
M30B (PSR~J2140-23B) is a 13-ms pulsar that we have only seen once
(but at very high significance in two pulsar back-ends) presumably due
to scintillation.  The pulsar is in a very eccentric orbit ($e>0.5$)
with an orbital period of 1$-$10\,days and a companion of minimum mass
$~0.35$\,\msun.  Future detections will allow the measurement of
$\dot\omega$ (and hence the total system mass) and probably
relativistic $\gamma$ which will provide the masses of both
companions.  More details on these two pulsars are available in
\cite{rsb+04}.

\subsection{M71}
M71 is another low mass and low density cluster; however, it is
relatively nearby at $\sim$4\,kpc.  We discovered the first (and so
far only) pulsar in this cluster in our initial observation of it with
Arecibo.  The pulsar is almost certainly a cluster member given its
location $\sim$0.6 core radii from the cluster center and a DM
($\sim$117\,pc\,cm$^{-3}$) that is reasonably close to the prediction
of the NE2001 Galactic electron model
\citep[$\sim$86\,pc\,cm$^{-3}$;][]{cl02}.  M71A is yet another
Black-Widow-like eclipsing binary (see fig.~\ref{fig:eclipsers}) with
$P_{psr}=4.89$\,ms and a minimum companion mass of $M_{c,\rm min} =
0.03$\,\msun.  Due to the relatively high DM, the pulsar does not
scintillate at 20\,cm, nor are delays in the pulse arrival times
visible at eclipse ingress or egress.  Its proximity and the low
density of the cluster (i.e. less source crowding) make M71A a good
target for both optical and X-ray observations.  More details will
soon be available in Stairs et al., in preparation.

\subsection{NGC~6749}
NGC~6749, which is at $\sim$7.9\,kpc, has the lowest concentration
($c=log({\rm r}_t / {\rm r}_c$)) and the second lowest central
luminosity density of any GC with known pulsars.  We have confirmed
one new MSP in NGC~6749, the 3.19-ms binary NGC~6749A.  This is the
first pulsar in this cluster, although we have a good but unconfirmed
candidate for NGC~6749B (4.97\,ms).  Timing observations are planned
with Arecibo and will allow us to solve the orbit of NGC~6749A,
hopefully confirm NGC~6749B, and potentially uncover additional
pulsars.

\begin{figure}
\plotone{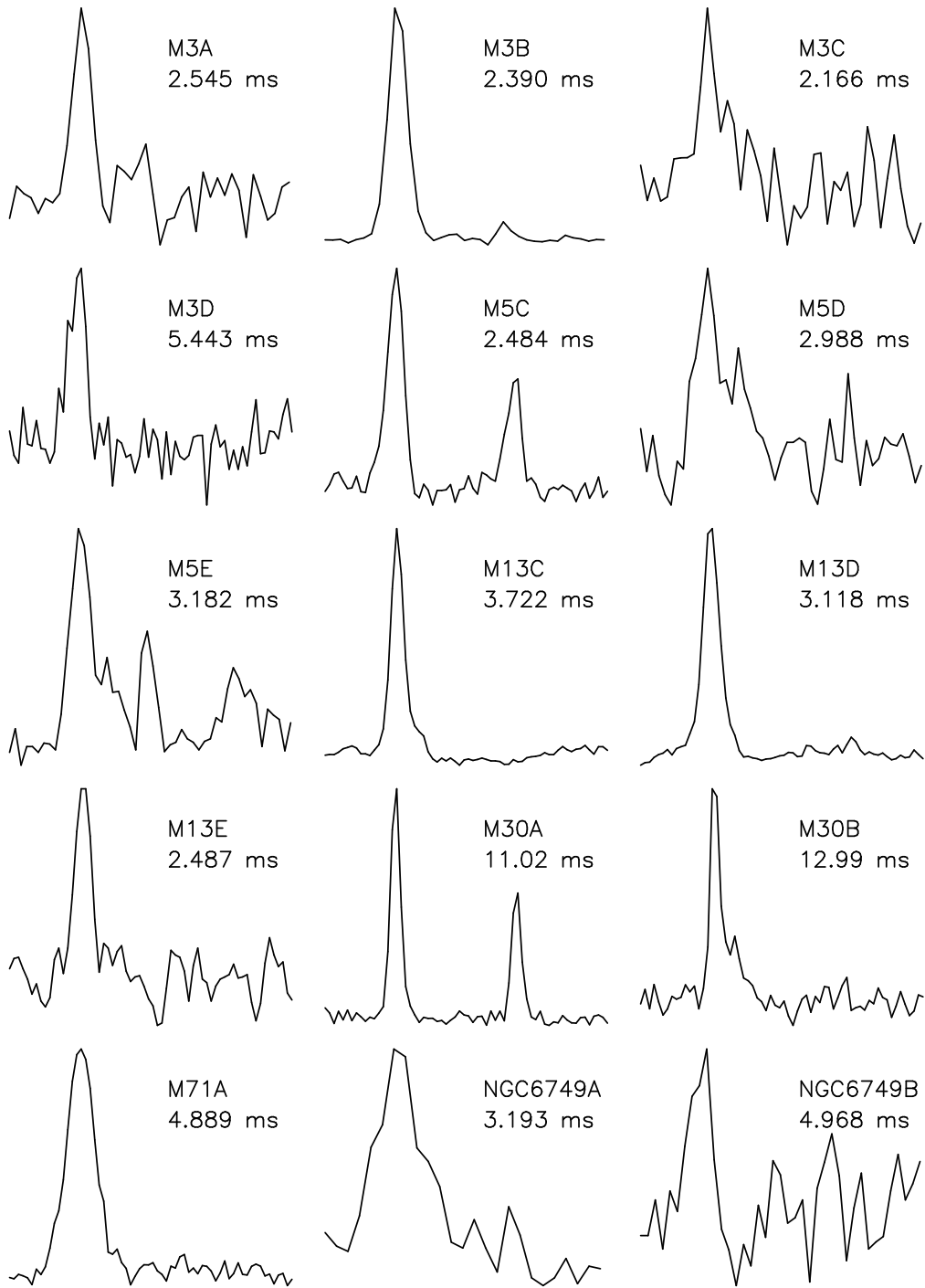}
\caption{Pulse profiles for 13 new MSPs and 2 candidates (M3C and NGC~6747B).\label{fig:profs}}
\end{figure}

\section{Conclusion}
Our on-going 20-cm survey of clusters using the GBT and Arecibo has
been very successful, resulting in the discovery of 13 new MSPs.
These detections have benefitted greatly from the upgraded Arecibo
telescope, the WAPP pulsar backends, and perhaps most importantly, a
prodigious amount of available computing power.  The vast majority of
these systems would not have been discovered without both
computationally expensive acceleration searches and repeated
observations of the clusters.  We anticipate determining timing
solutions for the majority of the systems --- or at least reliable
orbital parameters --- within the next year.  We also believe that
future multi-wavelength campaigns will uncover many more pulsars in
these
clusters.\\
{\em Acknowledgements:} We wish to thank the Canadian Foundation for
Innovation for the grant that purchased ``The Borg'' and made this
project possible.

%\bibliographystyle{apj} 
%\bibliography{apj-jour,pulsars,psrrefs}

\begin{thebibliography}{11}
\expandafter\ifx\csname natexlab\endcsname\relax\def\natexlab#1{#1}\fi

\bibitem[{Anderson(1992)}]{and92}
Anderson, S.~B. 1992, PhD thesis, California Institute of Technology

\bibitem[{{Anderson} {et~al.}(1997){Anderson}, {Wolszczan}, {Kulkarni}, \&
  {Prince}}]{awkp97}
{Anderson}, S.~B., {Wolszczan}, A., {Kulkarni}, S.~R., \& {Prince}, T.~A. 1997,
  \apj, 482, 870

\bibitem[{Backer {et~al.}(1997)Backer, Dexter, Zepka, Ng, Werthimer, Ray, \&
  Foster}]{bdz+97}
Backer, D.~C., Dexter, M.~R., Zepka, A., Ng, D., Werthimer, D.~J., Ray, P.~S.,
  \& Foster, R.~S. 1997, \pasp, 109, 61

\bibitem[{{Cordes} \& {Lazio}(2002)}]{cl02}
{Cordes}, J.~M. \& {Lazio}, T. J.~W. 2002, astro-ph/0207156

\bibitem[{{Dowd} {et~al.}(2000){Dowd}, {Sisk}, \& {Hagen}}]{dsh00}
{Dowd}, A., {Sisk}, W., \& {Hagen}, J. 2000, in ASP Conf. Ser. 202: IAU Colloq.
  177: Pulsar Astronomy - 2000 and Beyond, 275

\bibitem[{Harris(1996)}]{har96}
Harris, W.~E. 1996, \aj, 112, 1487

\bibitem[{Phinney(1992)}]{phi92b}
Phinney, E.~S. 1992, Phil. Trans. Roy. Soc. A, 341, 39

\bibitem[{{Ransom}(2001)}]{ran01}
{Ransom}, S.~M. 2001, PhD thesis, Harvard University

\bibitem[{{Ransom} {et~al.}(2003){Ransom}, {Cordes}, \& {Eikenberry}}]{rce03}
{Ransom}, S.~M., {Cordes}, J.~M., \& {Eikenberry}, S.~S. 2003, \apj, 589, 911

\bibitem[{{Ransom} {et~al.}(2002){Ransom}, {Eikenberry}, \&
  {Middleditch}}]{rem02}
{Ransom}, S.~M., {Eikenberry}, S.~S., \& {Middleditch}, J. 2002, \aj, 124, 1788

\bibitem[{{Ransom} {et~al.}(2004){Ransom}, {Stairs}, {Backer}, {Greenhill},
  {Bassa}, {Hessels}, \& {Kaspi}}]{rsb+04}
{Ransom}, S.~M., {Stairs}, I.~H., {Backer}, D.~C., {Greenhill}, L.~J., {Bassa},
  C.~G., {Hessels}, J.~W.~T., \& {Kaspi}, V.~M. 2004, \apj, 604, 328

\end{thebibliography}

\end{document}